\documentclass{aa}

\usepackage{graphicx}
\usepackage{txfonts}
\usepackage{wasysym}
\usepackage{color}

\begin{document} 

   \title{Rotation period distribution of CoRoT\thanks{The CoRoT space mission was developed and is operated by the
                     French space agency CNES, with the participation of ESA's RSSD and
                     Science Programmes, Austria, Belgium, Brazil, Germany, and Spain.} and {\em Kepler} Sun-like
          stars}

   \author{I.~C.~Le\~ao \inst{1}
   \and L.~Pasquini \inst{2}
   \and C.~E.~Ferreira Lopes \inst{1,3}
   \and V.~Neves \inst{1}
   \and A.~A.~R.~Valcarce \inst{4,5,6}
   \and L.~L.~A.~de~Oliveira \inst{1}
   \and D.~Freire~da~Silva \inst{1}
   \and D.~B.~de~Freitas \inst{1}
   \and B.~L.~Canto Martins \inst{1}
   \and E.~Janot-Pacheco \inst{7}
   \and A.~Baglin \inst{8}
   \and J.~R.~De~Medeiros \inst{1}
   }

\authorrunning{I. C. Le\~ao et al.}
\titlerunning{Rotation period distribution of CoRoT and {\em Kepler} Sun-like stars}

   \institute{Departamento de F\'isica, Universidade Federal do Rio Grande do Norte, Natal, RN, 59072-970 Brazil\\
              \email{izan@dfte.ufrn.br}
   \and ESO -- European Southern Observatory, Karl-Schwarzschild-Strasse 2, 85748 Garching bei M\"unchen, Germany
   \and SUPA (Scottish Universities Physics Alliance) Wide-Field Astronomy Unit, Institute for Astronomy, School of Physics and Astronomy, University of Edinburgh, Royal Observatory, Blackford Hill, Edinburgh EH9 3HJ, UK
   \and Pontificia Universidad Cat\'olica de Chile, Instituto de Astrof\'isica, Facultad de F\'isica, Av. Vicu\~na Mackena 4860, 782-0436 Macul, Santiago, Chile
   \and Pontificia Universidad Cat\'olica de Chile, Centro de Astroingenier\'ia, Av. Vicu\~na Mackena 4860, 782-0436 Macul, Santiago, Chile
   \and Millennium Institute of Astrophysics, Santiago, Chile
   \and Universidade de S\~ao Paulo/IAG-USP, rua do Mat\~ao, 1226, Cidade Universit\'aria, S\~ao Paulo, SP, 05508-900 Brazil
   \and LESIA, UMR 8109 CNRS, Observatoire de Paris, UVSQ, Universit\'e Paris-Diderot, 5 place J. Janssen, 92195 Meudon, France }

   \date{Received Month Day, Year; accepted Month Day, Year}

  \abstract
   {}
   {We study the distribution of the photometric rotation period ($P_{\rm rot}$),
    which is a direct measurement of the surface rotation at active latitudes,
    for three subsamples of Sun-like stars:
    one from CoRoT data and two from {\em Kepler} data.
    For this purpose, we identify the main populations of these samples and interpret their main biases
    specifically for a comparison with the solar $P_{\rm rot}$.}
   {$P_{\rm rot}$ and variability amplitude ($A$) measurements were obtained from public CoRoT and {\em Kepler} catalogs,
    which were combined with public data of physical parameters.
    Because these samples are subject to selection effects,
    we computed synthetic samples with simulated biases to compare with observations,
    particularly around the location of the Sun in the Hertzsprung-Russel
(HR) diagram.
    Publicly available theoretical grids and empirical relations were used to combine physical
    parameters with $P_{\rm rot}$ and $A$.
    Biases were simulated by performing cutoffs on the physical and rotational parameters in the same way as in each observed sample.
    A crucial cutoff is related with the detectability of the rotational modulation,
    which strongly depends on $A$.}
{The synthetic samples explain the observed $P_{\rm rot}$ distributions of Sun-like stars as having two main populations:
one of young objects (group~I, with ages younger than $\sim$ 1~Gyr)
and another of main-sequence and evolved stars (group~II, with ages older than $\sim$ 1~Gyr).
The proportions of groups~I and~II in relation to the total number of stars
range within 64--84\% and 16--36\%, respectively.
Hence, young objects abound in the distributions, producing the effect of observing a high number of short periods around
the location of the Sun in the HR diagram.
Differences in the $P_{\rm rot}$ distributions between the CoRoT and {\em Kepler} Sun-like samples
may be associated with different Galactic populations.
Overall, the synthetic distribution around the solar period agrees with observations,
which suggests that the solar rotation is normal with respect to Sun-like stars within the accuracy of current data.}
   {}


   \maketitle
%

\section{Introduction}
\label{intro}

A question of high interest is how exactly the Sun will evolve and, as a consequence, how it will affect the planetary environment.
Although the Sun's evolution has long been modeled from thermonuclear reaction rates,
explaining non-canonical effects, such as sunspots and magnetic fields, is still puzzling.
Clearly understanding the solar structure and evolution is necessary to know, for example, how long our central star will maintain the conditions required for life in the solar system, or whether its magnetic activity can damage our technological needs, among other concerns
\citep[e.g.,][]{mel11,jon13,shi13,ste13}.
To accurately predict the near and distant future characteristics of the Sun, it is essential to understand its past and project its future by studying other stars
\citep[e.g.,][]{cha10,cha13,eks12}.
Specifically, stars with a similar mass and chemistry to that of the Sun may provide important constraints for the theoretical models used in predicting the Sun's history
\citep[e.g.,][]{sch11,dat12}.
To refine these models, stellar rotation is fundamental because this parameter is a key feature in controlling the root-cause of the structure, evolution, chemistry, and magnetism of the stars \citep[e.g.,][]{pal13,mae14}.
In particular, the measure of
the surface rotation can substantially improve observational constraints on theory
from a proper modeling of the stellar interior \citep[e.g.,][]{zah92,eks12,bru14}.
Further refinements can be obtained from the internal rotation,
which has recently been measured in the Sun and several different types of stars
thanks to remarkable advances in asteroseismology \citep[e.g.,][]{gou13,deh14}.

Another key question is how similar or different the Sun is compared to Sun-like stars, and rotation may also be crucial in this comparison
\citep[e.g.,][]{gus98,gon99,gon01}.
Different studies have discussed whether the Sun rotates normally with respect to its Sun-like counterparts, at least considering the surface rotation. For example, \citet{sod83} and \citet{gra82} suggested that the Sun rotates normally for its age, a finding corroborated by \citet{rob08}. However, as reported by \citet{fre13}, these studies were based on the projected rotational velocity ($v\sin i$), which is an indirect rotation measurement with the intrinsically unknown axis orientation angle $i$.
In contrast, \citet{met13} found in a recent study of magnetic activity cycles of the solar analog $\epsilon$ Eri that this star undergoes long-term magnetic cycles compatible with those of the Sun, but the rotation period $P_{\rm rot}$ determined from the photometric time-series
\citep[e.g.,][]{str09}
is approximately twice as fast. Based on $P_{\rm rot}$, which is a direct measurement of the surface rotation at the active latitudes, it is suggested that the Sun may rotate more slowly than its analogs. Nevertheless, the question about solar rotation normality remains unanswered, especially when we consider its rotation compared to direct stellar rotation measurements.

To date, most comparative studies of the Sun's rotation relative to other stars are based on the projected rotational velocity from $v\sin i$
\citep[e.g.,][]{por97,mel06},
which, as explained, may hinder proper comparison. In contrast, the number of databases with direct $P_{\rm rot}$ measurements from the photometric time-series is growing rapidly
\citep[e.g.,][]{str00,str09,har10,med13,rei13,mcq14}.
Current examples of databases with stellar rotation periods are those obtained from the CoRoT\footnote{\tt http://smsc.cnes.fr/COROT/} \citep{bag09} and {\em Kepler}\footnote{\tt http://kepler.nasa.gov/} \citep{koc10} space telescope data, which include long-term light curves (LCs) with high-temporal resolution and high-photometric sensibility for more than 300,000 objects.
Together, these databases include tens of thousands of period measurements of suggested rotation or candidates
\citep[e.g.,][]{mei11,aff12,aff13,med13,nie13,mcq13a,mcq13b,mcq14,rei13,wal13}.
However, for many of these stars, no spectroscopic data are available. Thus, it is difficult to associate their periods with other physical parameters, such as metallicity, mass, radius, temperature, and age.

In particular, \citet{mcq13a,mcq14} found evidence of a bimodal $P_{\rm rot}$ distribution for cool stars that becomes shallower with higher temperatures.
Such a bimodality, also detected in \citet{rei13}, has been proposed to originate from stellar populations that evolved differently from one another.
The current lack of spectroscopic data is one of the main limitations in explaning this effect.
The {\em Kepler} database so far provides photometric estimates of physical parameters for nearly all targets that are useful for several purposes.
With regard to CoRoT, these photometric estimates are typically available for the targets of the asteroseismology channel.
Thus, follow-up is crucial to obtain this physical information for the targets of the exoplanet channel
(i.e., those considered in the present work).

Recently, \citet{gar14} examined the rotational evolution of solar-like stars based on the {\em Kepler} data.
These authors performed a refined study of the age-rotation-activity relations by combining
rotation period measurements with precise asteroseismic ages.
One of the results was an analysis of the evolution of the rotation period as a function of stellar age
for selected 12 cool dwarfs.
The period-age relation corresponded well with previous calibrations, such as the Skumanich law \citep{sku72}
and those obtained by~\citet{bar07} and~\citet{mam08}.
Based on the referenced work, a fit with the 12 selected stars is compatible with the Sun location in the period-age diagram.
However, the Sun lies at a slightly longer period than the fit, a small discrepancy that needs to be investigated in more detail.

For general studies of {\em Kepler} stars, \citet{hub14} provided a compilation of physical parameters obtained from different methods with improved values and well-computed errors by adjusting observational data with theoretical grids.
For CoRoT targets, \citet{sar13} recently published a relatively long list of physical parameters obtained automatically from spectroscopy. These data represent a unique opportunity for follow-up of the photometric periods determined from CoRoT and {\em Kepler} observations. In addition, a comparison between observations and theory can be performed by considering recent evolutionary models developed by the Geneva Team \citep{eks12}, which account for stellar rotation based on a rich set of physical ingredients (see Sect.~\ref{simul}). Accordingly, photometric periods can be compared with these models.
It is necessary for such a comparison to carefully analyze the observational bias of the observed sample, as described below.

Biases are typically modeled in stellar population synthesis, such as in the TRILEGAL\footnote{\tt http://stev.oapd.inaf.it/cgi-bin/trilegal} \citep[e.g.,][]{gir05} and SYCLIST\footnote{\tt http://obswww.unige.ch/Recherche/evoldb/index/} \citep{geo14} codes.
TRILEGAL combines nonrotating evolutionary tracks with several ingredients, such as the star formation rate, age-metallicity relation, initial mass function, and geometry of Galaxy components, to simulate, in particular, the distribution of physical parameters and photometric measurements of different stellar populations.
SYCLIST is a recent code developed by the Geneva Team that includes stellar rotation
in a detailed set of ingredients to compute synthetic parent samples.
The code considers anisotropies of the stellar surface,
such as the latitude-dependence of temperature and luminosity and their effects on the limb darkening,
to predict how the measured physical parameters can be affected by different rotation axis orientations.
The resulting biases can, for example, affect the age estimation
of coeval samples or of individual field stars because the physical parameters may deviate from their actual values.
Although the SYCLIST code can be used to predict deviations on rotation measurements that are the result of different axis orientations, it does not yet consider the detectability of $P_{\rm rot}$ measurements from LC variations.

For a deeper study of stellar rotation, understanding biases of $P_{\rm rot}$ measurements can be particularly useful.
As such, an important effect must be considered:
the amplitude of the rotational modulation decreases as long as the stars evolve \citep[e.g.,][]{mes01,mes03,gia11,spa11};
this decreases their detectability.
As a consequence, current field stellar samples with $P_{\rm rot}$ measurements
are biased, exhibiting a lack of stars around the
solar rotation period\footnote{The solar period ranges from 24.47 days at the equator to 33.5 days at the poles
and has an avearage of 26.09 days \citep{lan03}.}.
This bias hampers a proper analysis of the solar rotation normality regarding the $P_{\rm rot}$ distribution of Sun-like field stars.
Therefore, a quantified analysis is needed to better identify
these
biases in field stellar samples with variability measurements.

We here study the empirical distribution of $P_{\rm rot}$ for field stars with physical parameters similar to those of the Sun.
To this end, three different subsamples of Sun-like stars were obtained from the CoRoT and {\em Kepler} public catalogs,
and synthetic samples were built by reproducing the main biases of the observed samples.
Actual and synthetic samples can then be compared to one another to interpret the observational biases
and to study the solar rotation normality. This analysis is performed by testing the most recent theoretical predictions combined with a large amount of new observational data. Combining this observational and theoretical information in the same analysis allows us to conduct an unprecedented study concerning these CoRoT and {\em Kepler} targets compared with the solar rotation.

We begin by describing our methods (Sect.~\ref{methods}) for sample selection from the public CoRoT and {\em Kepler} data.
Next, we explain the methods we used to build the synthetic samples (Sect.~\ref{simul}).
We discuss the rotation period distribution of young stars (Sect.~\ref{pms}), which were needed to constrain our synthetic samples.
We present our results (Sect.~\ref{results}) starting from the validation between the observational data and theoretical predictions for the Sun-like CoRoT sample. This sample, obtained from \citet[hereafter DM13]{med13} was carefully considered as a catalog of rotating candidates because
physical information of its sources was very limited during its development.
Now, new spectroscopic physical parameters provided by \citet{sar13} allow us to
determine whether this sample is indeed a catalog of $P_{\rm rot}$. Finally, the $P_{\rm rot}$ distribution is analyzed for the CoRoT and {\em Kepler} subsamples of Sun-like stars by interpreting biases and by comparing the stars with the solar rotation.


\section{Sample selection and physical parameters}
\label{methods}

In this section, we summarize the catalogs used to obtain
our observational samples, from which we constructed the corresponding synthetic samples.
These include the CoRoT and {\em Kepler} catalogs with variability period measurements
and those used to derive the physical parameters.

\subsection{Selection of CoRoT targets and their parameters}


From a parent sample of $124,\!471$ LCs,
DM13 presented a catalog of $4,\!206$ CoRoT stellar rotating candidates with unambiguous variability periods. A detailed procedure was developed for proper selection of the targets, based only on CoRoT LCs and 2MASS photometry, before additional data such as spectroscopy data were available. The procedure includes an automatic pre-selection of LCs that corrects them for jumps, long-term trends, and outliers, and identifies a preliminary list of periods, amplitudes ($A$), and signal-to-noise ratios ($S/N$). The variability period was computed from Lomb-Scargle \citep{lom76,sca82,hor86}
periodograms, the amplitude from a harmonic fit of the phase diagram for the main period, and the $S/N$ was the amplitude-to-noise ratio calculated from the LC. These parameters were then used to select the highest quality LCs (high $S/N$) in the expected amplitude and period ranges for rotating candidates. This criterion is related to the detectability of rotational modulation in an LC and was used in the present work as a relevant ingredient to produce a synthetic sample with biases (see Sect.~\ref{simul}). Finally, visual inspection was performed by applying a list of well-defined criteria for identifying photometric variability compatible with the rotational modulation caused by dynamic star spots (semi-sinusoidal variation\footnote{Semi-sinusoidal variability is characterized by some asymmetry of the maximum and minimum flux with respect to its average over time, somewhat irregular long-term amplitude variations, and semi-regular multi-sinusoidal short-term flux variations, typically with an amplitude $\lesssim$~0.5~mag and period $\gtrsim$~0.3~days. This description is based on the dynamic behavior of star spots, as observed for the Sun (see DM13, Sect.~2.2).} defined in DM13), and the final parameters were refined.


\citet{sar13} used neural networks similar to those employed in the CoRoT Variable Classifier (CVC) tool \citep{deb07,deb09,blo10} to obtain physical parameters from spectroscopic data of $6,\!832$ CoRoT targets. The method was applied to FLAMES/GIRAFFE\footnote{\tt http://www.eso.org/sci/facilities/paranal /instruments/flames/} spectra by using two different types of training sets. One is composed of synthetic spectra from Kurucz models \citep{ber08} and TLUSTY grids \citep{hub95} (KT set) and the other of actual ELODIE\footnote{\tt http://www.obs-hp.fr/guide/elodie/elodie-eng.html} spectra with known parameters from the ELODIE library \citep{pru01,pru04} (ELODIE set). Physical parameters, particularly gravity ($\log g$) and effective temperature ($T_{\rm eff}$), were then obtained independently for both sets. The main limitations of this catalog are the uncertainties in the computed parameters. Thus, no metallicity values are available and the errors in $\log g$ and $T_{\rm eff}$ are relatively high (originally with typical values of approximately 0.5~dex and 400~K, respectively). As a first step to optimize our analysis despite the limitations, we averaged multiple independent estimations of $\log g$ and $T_{\rm eff}$ provided by \citet{sar13} for each target, which decreased the typical uncertainties slightly to approximately 0.4~dex and 300~K, respectively.

For the sample selection, we combined the catalog of DM13 with the available physical parameters given by \citet{sar13}. This provided a subsample of 671 targets, namely the DMS sample, which we analyze here. Because \citet{sar13} did not provide metallicity values, we assumed that the sample has a mixture of metallicities with an average around the solar value. This assumption is reasonable given that the overall metallicity $[M/H]$ distribution in the CoRoT fields, in accordance with \citet{gaz10},
has a mean value and standard deviation of $-0.05 \pm 0.35$~dex.
For a particular analysis of stars with
$T_{\rm eff}$ and $\log g$ similar to the solar values,
we selected a small subsample of 175~stars, namely, the DMS$_{\odot}$ sample, within an elliptical region
of the size of typical uncertainties
around $T_{\rm eff} = 5772 \pm 300$~K and $\log g = 4.44 \pm 0.4$~dex.
Then, we studied the distribution of $P_{\rm rot}$ for this sample.
Despite the large errors in $\log g$ and $T_{\rm eff}$ and the assumptions on metallicity, the DMS$_{\odot}$ sample can be interpreted
from a comparison with a simulated sample, which considers the main observational
biases and uncertainties, as described in Sect.~\ref{simul}.

\subsection{\it Selection of {\em Kepler} targets and their parameters}

Several recent works \citep{mcq14,mcq13a,mcq13b,nie13,wal13,rei13} provide
rotation periods for a large portion of the {\em Kepler} targets.
In particular, \citet{rei13} derived periods for $24,\!124$ targets
observed during quarter Q3 by following
an automatic method with partial inspection.
In the method, the range between the 5th and 95th percentile of the normalized flux distribution of a LC,
named {\it \textup{variability range}} $R_{\rm var}$ \citep{bas10,bas11},
is an adimensional quantity with a numerical value approximately equal to the variability amplitude $A$ in units of magnitude.
The detectability of rotating stars was primarily determined by defining an $R_{\rm var}$ threshold (of $3\permil$),
which produced a bias conceptually comparable to the $S/N$ threshold of DM13.
The catalog of \citet{rei13} provides rotation periods ranging from 0.5 to 45 days,
thus covering a region that can be compared with the solar rotation period.

More recently, \citet{mcq14} provided a new catalog of rotation periods ranging from 0.2 to 70~days for $34,\!040$ {\em Kepler} targets
observed during quarters Q3--Q14 (which is a much longer time span than considered in \citealt{rei13}).
This is currently the largest {\em Kepler} sample of rotation periods and was obtained from a fully automatic method named AutoACF.
The method is based on the autocorrelation function, which was applied to a trained neural network that selected a sample of stars exhibiting rotational modulation.
For the period detectability, the authors defined a weight $w$ that has a sophisticated calculation
based on the ACF periodogram, physical parameters, and neural network selection.
This weight, with a threshold $w_{\rm thres} = 0.25$,
produces a bias in the detectability of the rotational modulation,
somewhat similar to the aforementioned $S/N$ or $R_{\rm var}$ parameters.

\citet{hub14} recently presented a catalog of physical parameters for {\em Kepler} stars
that can be combined with rotation period measurements.
These authors provided a large compilation of physical parameters estimated from different methods for $196,\!468$ stars observed in the {\em Kepler} quarters 1 to 16. For the majority of the targets, atmospheric properties (temperature, gravity, and metallicity) were estimated from photometry and have high uncertainties. In addition, there is also a noticeable number of objects with more refined measurements obtained from spectroscopy, asteroseismology, and exoplanet transits. Finally, the authors refined the parameters by adjusting observations with theoretical grids. For this reason, this catalog has a sharp distribution in the Hertzsprung-Russel
(HR) diagram, as shown in Sect.~\ref{simul}, which is also suitable for our methods used to produce a synthetic sample.
On average, the typical uncertainties in $T_{\rm eff}$, $log g$, and metallicity $[Fe/H]$ lie in the refereed catalog at approximately 170~K, 0.2~dex, and 0.2~dex, respectively.
To obtain a subsample of {\em Kepler} Sun-like stars,
we selected only targets with metallicity $-0.2$~dex~$< [Fe/H] < 0.2$~dex and defined an elliptical region in the HR~diagram within $T_{\rm eff} = 5772 \pm 170$~K and $\log g = 4.44 \pm 0.2$~dex.
This region was used to select Sun-like stars in the catalogs of \citet{rei13} and \citet{mcq14}, each combined with \citet{hub14},
namely, the RH$_{\odot}$ and MH$_{\odot}$ samples, which
are composed of $1,\!836$ and $2,\!525$ objects, respectively.
The same metallicity range and HR~diagram region was used to produce the synthetic forms of these two {\em Kepler} Sun-like samples.

\section{Synthetic sample and main biases}
\label{simul}


\citet{eks12} presented a new version of the Geneva stellar evolution code \citep{egg08}
for noninteracting stars that includes theoretical predictions for the surface rotation.
To predict this parameter, the authors considered several physical assumptions regarding the stellar interior.
The main basis of this model is the shellular-rotation hypothesis \citep{zah92},
which states that the turbulence is substantially stronger in the horizontal than in
the vertical direction, yielding nearly constant angular rotation in each isobaric region.
On this basis, horizontal diffusion is combined with meridional circulation
and shear turbulence to predict transport mechanisms of angular momentum
and of chemical species.
The angular momentum is conserved during the whole stellar evolution, considering the mass-loss influence.
Finally, both atomic diffusion and magnetic braking in low-mass stars are accounted for in a homogeneous fashion.
The model provided grids of evolutions of solar-metallicity stars for masses ranging from 0.8 to 120~M$_{\odot}$, with low-mass stars evolving from the zero-age main sequence (ZAMS) to the core helium-burning phase.
Therefore, these grids provide a theoretical distribution of $P_{\rm rot}$ without observational biases.
From this theoretical distribution, synthetic biases and uncertainties
can be applied to study how these effects may modify actual observations.
Our analysis aims to resolve the actual $P_{\rm rot}$ distributions of three Sun-like samples
(the DMS$_{\odot}$, RH$_{\odot}$, and MH$_{\odot}$ sample) by an inversion procedure,
where we used the theoretical grids of \citet{eks12} as a root ingredient
to compute the three respective synthetic samples.
For these calculations, we identified the main biases
and defined appropriate parameters to describe each $P_{\rm rot}$ distribution.
These parameters were then properly adjusted to fit the synthetic distributions with the observations.

In summary, a synthetic parent sample of solar metallicity, obtained from an actual parent sample (see details below),
was generated with the parameters $T_{\rm eff}$ and $\log g$ assumed to be error-free.
These data were used as input to add the following parameters to the parent sample:
$P_{\rm rot}$, $A$, $S/N$, and $w$, also assumed to be error-free.
These parameters were set based on a combination of ingredients --
such as theoretical grids of \citet{eks12}, expected $P_{\rm rot}$ evolution of young stars from \citet{gal13}, and
$P_{\rm rot}$ \textit{versus} $A$ empirical relations from \citet{mes01,mes03} -- as detailed below.
Next, random fluctuations were applied to all the input and output parameters according to their errors in the actual sample.
From this step, only virtual targets with a certain detectability parameter greater than a given threshold were selected.
This parameter was defined according to each Sun-like sample
to reproduce potential biases originating from the instrumental capability of detecting rotational modulation.
For the DMS$_{\odot}$ sample, the $S/N$ was used to perform a cutoff similar to that stated in~DM13.
For the RH$_{\odot}$ sample, $A$ was used to mimic the selection by $R_{\rm var}$ performed in \citet{rei13}.
For the MH$_{\odot}$ sample, an empirical relation between $A$ and $w$ was identified and used to generate random synthetic values
of $w$ from the previously drawn $A$ values.
For every synthetic sample, we tested the bias produced by a cutoff with different parameters, $S/N$, $A$, and $w$, to discuss the effects on the $P_{\rm rot}$ distribution (see Sect.~\ref{results}).
Finally, only stars with properties similar to the Sun were selected by applying the same boundaries as for each of the three actual samples.
Because our final samples have physical parameters similar to the solar values,
the synthetic parent samples did not need to simulate stars well
that are very different from the Sun.

To further elaborate, the synthetic sample was
based on the following assumptions:

\begin{enumerate}[i.]

\item The parent distribution of each Sun-like sample is similar to the sample of \citet{sar13} for CoRoT targets or to that of \citet{hub14} for {\em Kepler} targets.

\item The rotation period follows a spread distribution for stars younger than $\sim$1~Gyr
      and converges to a common evolution per stellar mass, as proposed by \citet[][see their Fig.~3]{gal13}.

\item Based on the latter assumption, the synthetic parent sample is divided into two groups:
      one of young stars (group~I), with ages younger than $\sim$1~Gyr, and another of MS and evolved stars (group~II),
      with ages greater than $\sim$1~Gyr.
      For group~I, the actual period distribution is difficult to determine accurately (see Sect.~\ref{pms});
      thus, random values can be set within a modeled distribution to analyze how they may be superimposed with group~II.
      For group~II, the period follows the theoretical predictions provided by \citet{eks12}.

\item The detectability of rotational modulation depends on a certain parameter threshold:
      $S/N$ for the DMS$_{\odot}$ sample, $A$ for the RH$_{\odot}$ sample,
      and $w$ for the MH$_{\odot}$ sample.

\item The detectability parameter is strongly correlated with
      the variability amplitude or is the amplitude itself, which is related to the rotation period,
      as reported by \citet{mes01,mes03}.
      
\item The final $P_{\rm rot}$ distribution is also affected by uncertainties
      in $T_{\rm eff}$ and $\log g$, which cause the sample to become more widely distributed.

\end{enumerate}

\begin{figure}
\begin{flushright}
\includegraphics[width=8.8cm]{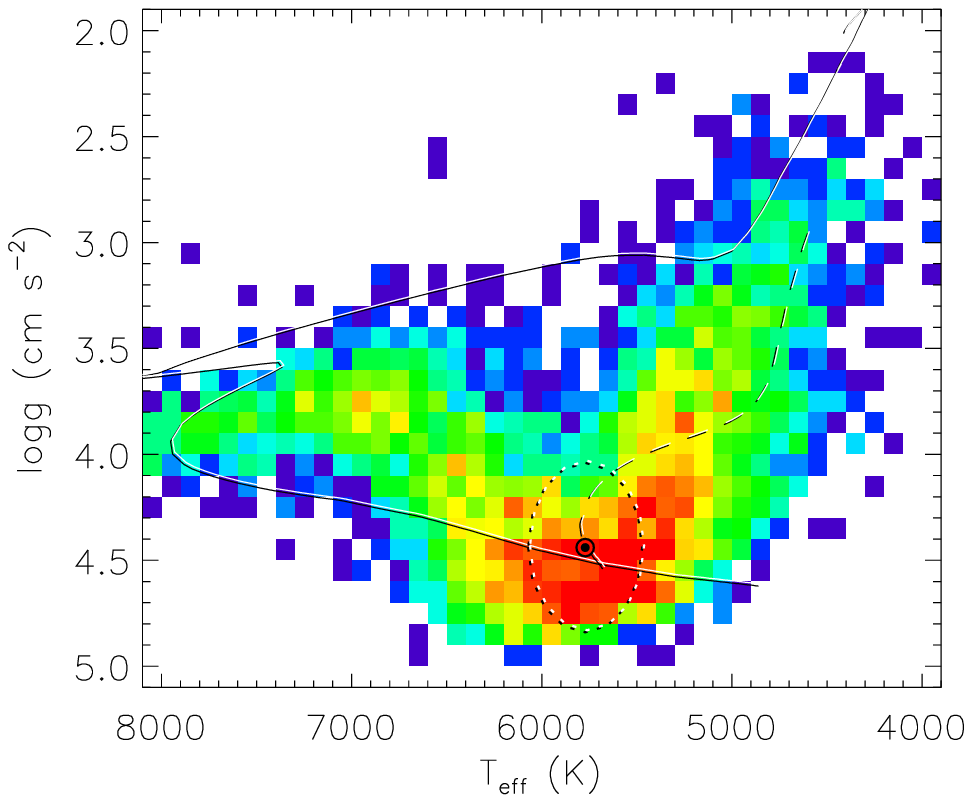}
\includegraphics[width=8.8cm]{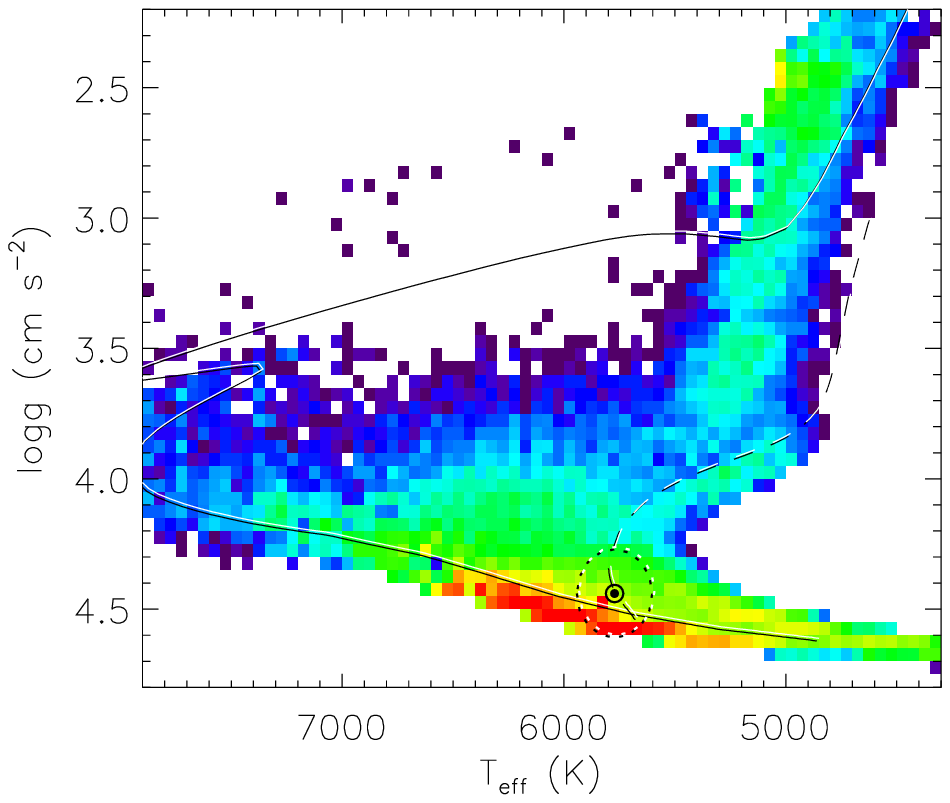}
\caption{
Probability distribution in the HR diagram for the parent samples considered in this work. {\it Top panel:} distribution for the CoRoT field, obtained from the full sample of \citet{sar13}. {\it Bottom panel:} distribution for the {\em Kepler} field, computed from a subselection of \citet{hub14} with a metallicity $-0.2$~dex~$< [Fe/H] < 0.2$~dex.
The red--green--blue color gradient in the map represents decreasing probability; the high end is the red part of the gradient.
Solid lines indicate the 1~Gyr isochrone of \citet{eks12}, and dashed lines illustrate the 1~M$_{\odot}$ theoretical track provided by the same authors. The Sun is represented by its standard symbol. The ellipses around the solar location depict the subsamples of Sun-like stars (Sun-like samples). These were selected equally
within the typical uncertainties in $T_{\rm eff}$ and $\log g$ for each parent sample for observed and synthetic data (see text).
}
\label{hessparent}
\end{flushright}
\end{figure}

To reproduce the parent distribution (as listed above in item~i), we first computed a Hess-like HR~diagram of
the original sample of \citet{sar13} for the CoRoT targets or a subsample of \citet{hub14} with metallicity $-0.2$~dex~$< [Fe/H] < 0.2$~dex for the {\em Kepler} targets.
These distributions were then used as a map of probabilities, as shown in Fig.~\ref{hessparent}.
The map resolution of the {\em Kepler} sample is higher than that of the CoRoT sample because of the considerably greater number of observed {\em Kepler} targets compared with CoRoT targets.
Therefore, a more refined study can be developed from the {\em Kepler} samples.
Actual samples were used as parent samples instead of population syntheses
because our methods aim to reproduce the detectability bias of the rotational modulation
and do not account for how the parent population was produced.
As such, all biases related to the observed fields are already implicit in the considered parent samples.
Population syntheses can also be used in this method if desired.
As another possibility, an image deconvolution could be applied to the parent sample probability map
to reduce the error spread. We tested this potential effect using a Richardson-Lucy algorithm (\citealt{ric72,luc74}; see also \citealt{lea06})
and obtained similar results to those shown in Sect.~\ref{results} with some fluctuations.
This suggests a robustness of the method, which produces stable results for different modelings of the parent distribution.

From each of the maps depicted in Fig.~\ref{hessparent}, a synthetic parent sample with $10^7$~stars
was generated randomly with probabilities respecting the map distribution.
The drawn values of $T_{\rm eff}$ and $\log g$ were then defined as being exact for the synthetic sample.
Then, the generated sample was plotted with the theoretical grid of \citet{eks12} to
perform the following definitions:

\begin{enumerate}[i.]

\item For $T_{\rm eff}$ out from the grid domain, the nearest valid value was set for a synthetic star.

\item Synthetic stars generated below the 1~Gyr isochrone
 of \citet{eks12} were defined to belong to group~I,
the remaining stars were designated to compose group~II.

\end{enumerate}
This means that all stars generated from the parent sample probability maps were used, including those drawn out from the grid domain.

When groups~I and~II were established,
theoretical predictions and empirical relations were used to aggregate $P_{\rm rot}$, $A$, and $S/N$
values to the synthetic parent sample.
First, random periods were set to the synthetic stars of group~I by following a custom distribution.
The empirical distribution of this region is discussed in Sect.~\ref{pms} to propose a synthetic distribution that may be superimposed with group~II.
For the synthetic stars of group~II, their $T_{\rm eff}$ and $\log g$ values (initially defined as being exact)
were used to set them to the corresponding rotation periods given in the \citet{eks12} theoretical grid.
The grid, valid for the solar metallicity, was built by interpolating theoretical tracks and isochrones altogether.

To set up $A$ values for the synthetic sample,
we considered the empirical relations of $P_{\rm rot}$ \textit{versus} $A$
obtained by \citet{mes01,mes03}, which were sorted according
to spectral type.
In these relations, the stars are somewhat uniformly distributed below their corresponding $A(P_{\rm rot})$ empirical functions.
The actual unbiased distribution in the $P_{\rm rot}$ \textit{versus} $A$ diagram cannot be derived from observations because observed data are biased.
Such a distribution could be obtained from theoretical models if spot-induced variability amplitude calculations
were included in the physical ingredients; however, such calculations were not implemented in the current models.
For the synthetic stars, $A$ values were therefore set up as uniform random values below $A(P_{\rm rot})$ curves
\citep[provided in][]{mes01,mes03}.
These curves were used for different spectral types, which were defined according to the
synthetic values of $T_{\rm eff}$ \citep[see][]{hab81}:
M-type for $T_{\rm eff} \leq 3700$~K, K-type for $3700$~K~$< T_{\rm eff} \leq 5200$~K, and G-type for $T_{\rm eff} > 5200$~K.

The approximation of considering a uniform probability distribution below the empirical relations of \citet{mes01,mes03}
may affect the final period distribution, especially at short periods, where the period-amplitude diagram is more widely distributed.
One main effect in the simulations is a change in the proportions of groups~I and~II
with respect to the total number of stars, $\rho_I$ and $\rho_{II}$, respectively.
These proportions are also sensitive to the line that defines
these groups (here the 1~Gyr isochrone),
which may fluctuate somewhat around its assumed HR diagram region
and may have systematics.
To adjust these limitations in the present work, the ratio $\rho_I/\rho_{II}$
was therefore set as a free parameter.
Adjusting these proportions
might be thought to produce a degeneracy:
the synthetic $P_{\rm rot}$ distribution would fit the observations
without the need of a detectability threshold such as $S/N$, $A,$ or $w$.
We checked that combining this adjustment with a proper detectability threshold
is certainly important to obtain an optimal fit,
otherwise, a noticeable discrepancy can occur, as explained in Sect.~\ref{mcq14}.
For the CoRoT sample, a detectability threshold is mandatory to obtain an acceptable fit while it has a weaker effect on the {\em Kepler} samples, but it does contribute to improve the fits.
Still, the $P_{\rm rot}$ distribution is bimodal regardless of its detectability bias.

\begin{figure}
\begin{flushright}
\includegraphics[width=8.8cm]{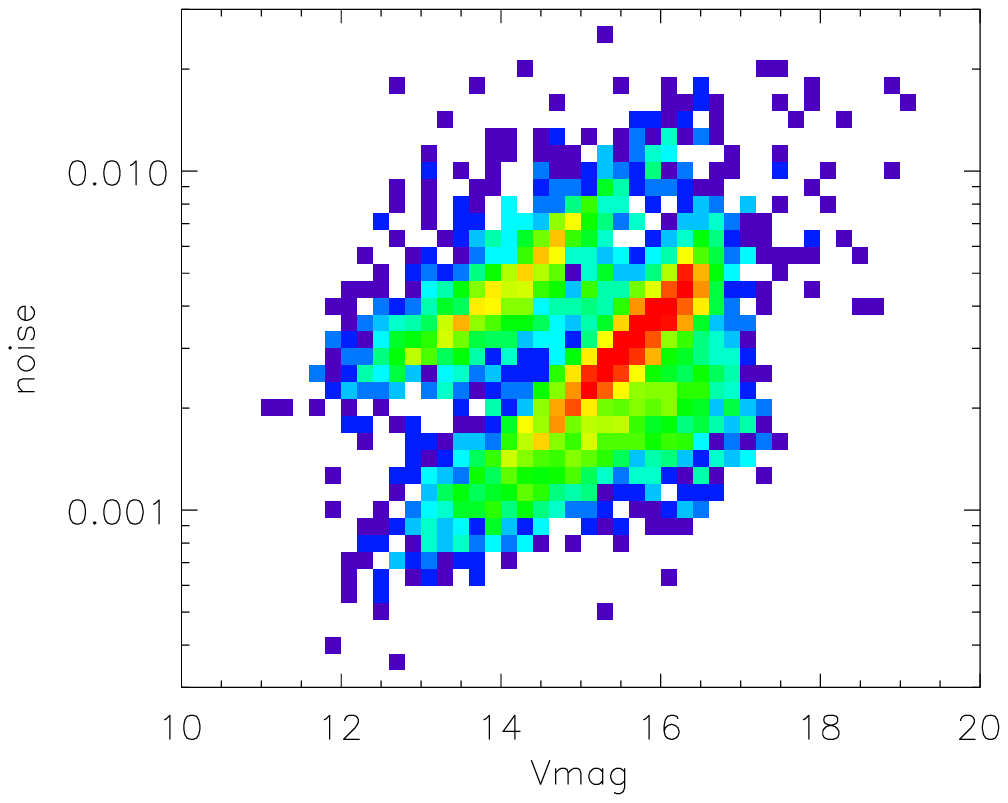}
\includegraphics[width=8.8cm]{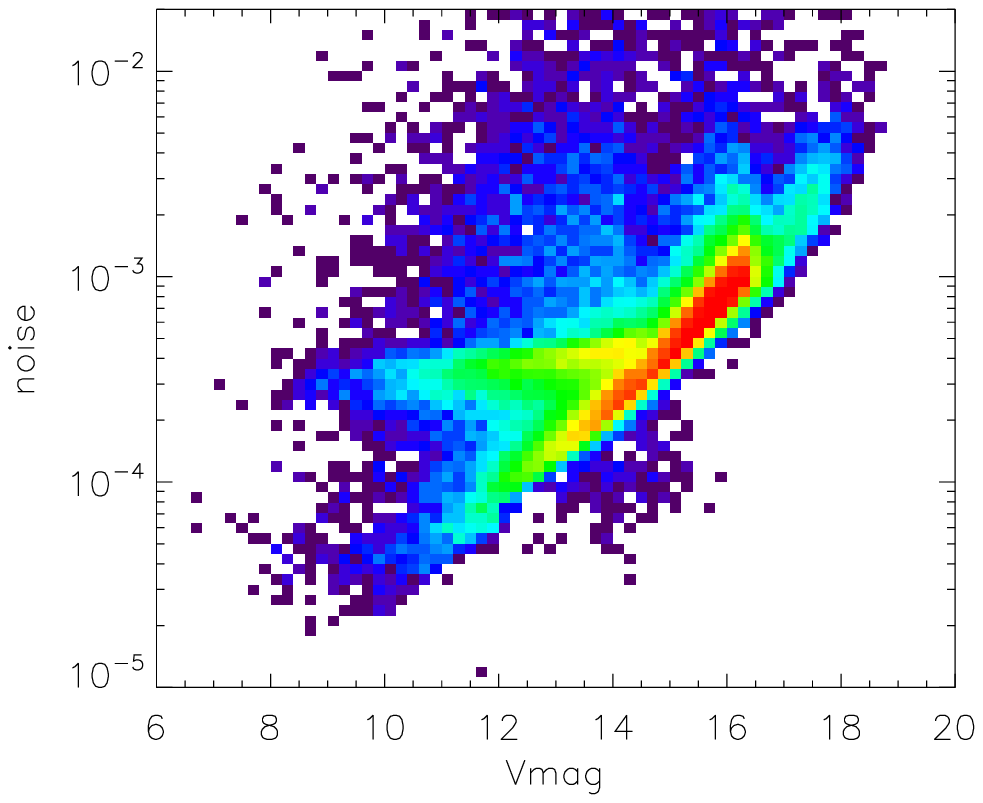}
\caption{
Probability maps for the empirical magnitude-noise diagrams of the CoRoT (top panel) and {\em Kepler} (bottom panel) parent samples.
}
\label{magnoise}
\end{flushright}
\end{figure}

Furthermore, empirical magnitude-noise diagrams were built from actual data: one from CoRoT data and another from {\em Kepler} data (see Fig.~\ref{magnoise}).
These maps were used to generate random magnitudes and noise levels for each corresponding synthetic sample.
We estimated the noise level based on Eq.~(1) given by~DM13.
This is a simple filtering of the high-frequency contribution that approximates the actual noise when the variability signal has a period considerably longer than the LC cadence.
The noise level estimated from this equation has the advantage of being simplistic because no prior fit is needed for the calculation. We verified that this noise level exhibits a similar behavior as a noise level
estimated from the residual of a harmonic fit or a boxcar smooth of the LC.
The two maps shown in Fig.~\ref{magnoise} have two branches each, which are explained
by the different types of LC bins (astero- versus exo-fields for CoRoT or short- versus long-cadence for {\em Kepler}).
Once noise levels were set to all synthetic stars,
the amplitudes drawn above were used to set reasonable
$S/N$ values for the synthetic sample.
In principle, $S/N$ is not needed to simulate the {\em Kepler} samples, but it was used
to test how the $P_{\rm rot}$ distribution depends on the cutoffs of different detectability parameters
(see Sect.~\ref{mcq14}).

\begin{figure}
\begin{flushright}
\includegraphics[width=8.8cm]{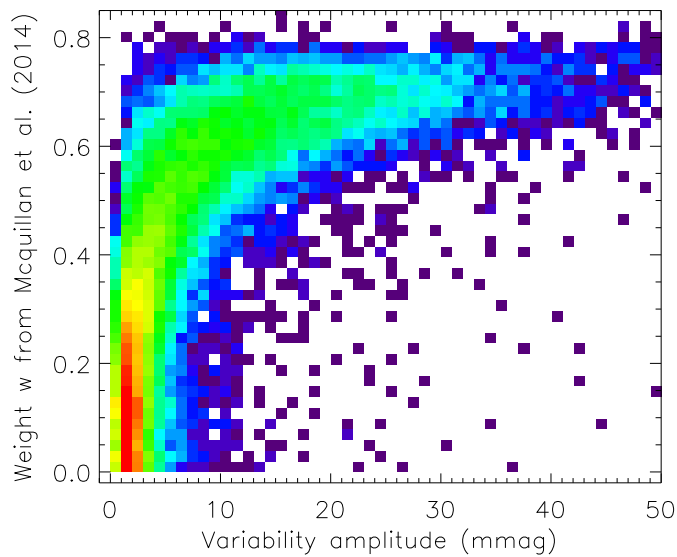}
\caption{
Probability map for the empirical amplitude-weight diagram, $A$ versus $w$, of the {\em Kepler} parent sample.
Public $w$ values were obtained from \citet{mcq14}, but we independently
computed the $A$ values.
}
\label{weiamp}
\end{flushright}
\end{figure}

Generating $w$ values, which are needed to simulate the MH$_{\odot}$ sample,
is particularly challenging because it is sophisticated.
For an approximate solution, we considered an empirical distribution
of $A$ versus $w$, which was used as a probability map (see Fig.~\ref{weiamp}).
In the public data of \citet{mcq14}, $w$ values are available for the entire parent sample,
but amplitudes are only given for $w > 0.25$.
Hence, we computed independent $A$ values from all {\em Kepler} LCs of the parent sample.
This was performed by automatically combining multiple quarters, as in~\citet{mcq14},
detecting the main periods of Lomb-Scargle periodograms and
computing the amplitude of their corresponding phase diagrams, regardless of their variability natures.
This simple procedure produced amplitude estimates that were highly compatible with those provided in~\citet{mcq14}.
From the probability map of Fig.~\ref{weiamp}, each synthetic $w$ value was obtained by respecting the probability distribution
of a section of the map according to a synthetic $A$ value drawn previously.

Next, random fluctuations were applied to the synthetic generated values of $T_{\rm eff}$, $\log g$, $P_{\rm rot}$, $A$, and $S/N$ according to their observational errors (no error was established for $w$).
Then, a synthetic subsample with $S/N$, $A,$ or $w$ values greater than a threshold was selected as being a biased sample, as in \citet{med13}, \citet{rei13}, and \citet{mcq14}.
These thresholds were computed as free parameters to obtain the best fits between the synthetic and the observed $P_{\rm rot}$ distributions.
Finally, an elliptical region around the solar $T_{\rm eff}$ and $\log g$ values was defined according to their typical uncertainties,
and a synthetic subsample of Sun-like stars was obtained.
A particular cutoff of $P_{\rm rot} < 45$~d was also performed for the RH$_{\odot}$ sample
and another of $T_{\rm eff} < 6500$~K was applied for the MH$_{\odot}$ sample to mimic similar cutoffs in the observed data, as performed in the original works.
Best fits between synthetic and actual samples were computed from the Kolmogorov-Smirnov (K-S) test
by minimizing the distance between their cumulative distributions.
Uncertainties were estimated by analyzing the parameter ranges in which the K-S distance was smaller than twice its minimum.

\section{Period distribution of young stars}
\label{pms}

\citet{gal13} developed models to predict the rotational evolution for pre-main sequence (PMS) stars.
In these models, stars with different initial angular velocity values evolve by keeping this parameter constant during the disk accretion phase. The angular velocity experiences variation as long as the radiative core evolves and decouples from the convective envelope. In this process, all stars reach a common evolution after $\sim$1 Gyr.
In our approach, we therefore propose to separate a synthetic parent sample into groups I and II (see Sect.~\ref{simul}),
with ages younger and older than 1~Gyr, respectively.
If the theoretical (unbiased) distribution for group I
are known, then observational biases can be implemented artificially to produce our synthetic sample.
However, such an unbiased distribution
is difficult to determine. We therefore discuss possible models that can be superimposed with group~II.

\begin{figure}
\begin{flushright}
\includegraphics[width=8.8cm]{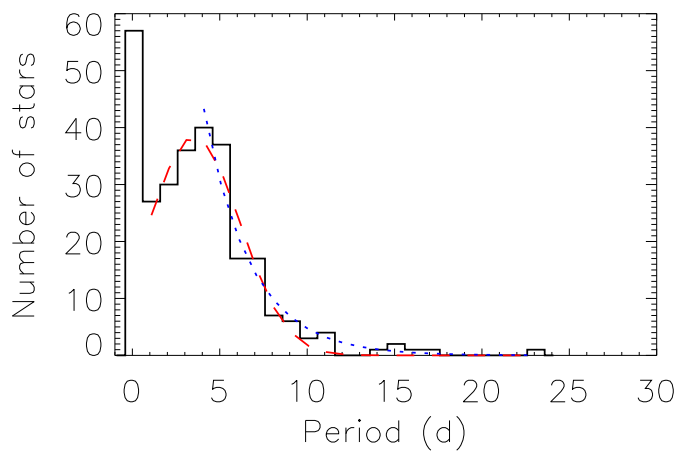}
\caption{
Rotation period distribution of the PMS Sun-like cluster stars selected by \citet{gal13}.
These stars belong to 13 star-forming regions or open clusters with ages ranging from 1~Myr to 1~Gyr and masses between 0.9 and 1.1~M$_{\odot}$.
The red dashed curve is a Gaussian fit with this distribution for periods greater than one~day (see text).
The blue dotted curve is an exponential fit with this distribution,
where the origin of the exponential function was placed in the distribution peak (see text).
}
\label{histgallet}
\end{flushright}
\end{figure}

Figure~\ref{histgallet} presents a histogram with the empirical distribution of rotation periods for a compilation of PMS Sun-like rotating stars obtained by~\citet{gal13}.
The stars belong to 13 star-forming regions or open clusters with ages between 1~Myr and 1~Gyr
and were selected within a mass range of 0.9 to 1.1~M$_{\odot}$.
The figure also shows two fits with simple models: a Gaussian function and an exponential function.
These fits were defined from two basic assumptions.
First, the empirical distribution from~\citet{gal13} can be basically interpreted as having two populations: one of fast rotators, with periods shorter than $\sim$1~d, and another with a spread of rotation periods peaking at approximately 4~days and ranging from 1 to 23~days.
Second, the interest for a comparison with the solar rotation period lies in the region of decreasing probability (to the right of the peak), which is the part that can be superimposed with group~II.

Therefore, the Gaussian fit was obtained within the domain where $P_{\rm rot} > 1$~d, that is, excluding the fast rotators.
The fit peaks at approximately 3.5~d and has a standard deviation $\sigma \simeq 2.6$~d.
A group of fast rotators was not identified in the CoRoT and {\em Kepler} Sun-like samples (see Sect.~\ref{results});
thus, for our applications, no cutoff was needed at short periods to compute Gaussian fits.
This most likely occurred because the period range of fast rotators coincides with many
other phenomena, such as pulsation and binarity; thus, more data than simply photometry data are needed to distinguish variability signals
within this range.
The exponential fit was designated as an approximation for the region of decreasing probability, in this case, for $P_{\rm rot} > 4$~d. This fit, defined with an origin at 4~d, has an exponential decay rate of approximately $-0.4$~d$^{-1}$.

Because the sample of~\citet{gal13} is a compilation of different works, a complex mixture of biases certainly affected this distribution.
As a consequence, the precise unbiased distribution of group~I is difficult to determine empirically.
Despite these facts, we represent the distribution of group~I by the two functions considered in the fits to test their superpositions with group~II.
For each sample analyzed in the results (Sect.~\ref{results}),
different Gaussian and exponential functions were tested with free parameters to fit
the corresponding data, which helped us to interpret the full $P_{\rm rot}$ distributions.


\section{Results and discussion}
\label{results}

In this section, we present the main results obtained from the subsamples considered in this work.
For the CoRoT sample, we first present (Sect.~\ref{corothr}) an evolutionary study of the full DMS sample
before analyzing the $P_{\rm rot}$ distribution of the DMS$_{\odot}$ sample.
This study is useful for validating the period measurements of DM13 as being rotational
and allows for an initial interpretation of the $P_{\rm rot}$ empirical distribution in the HR diagram.
In particular, there is a clear deviation between the observed $P_{\rm rot}$ average of Sun-like stars
and the solar value that is caused by biases (see below). Explaining this effect is the main motivation of this work.

\begin{table}
   \caption{Best-fit solutions.}
   {\centering
   \begin{tabular}{c c c c}
   \hline\hline
    ~ & \multicolumn{3}{ c}{Sample name} \\
     Parameter $^a$ & DMS$_{\odot}$ & MH$_{\odot}$$^b$ & RH$_{\odot}$ \\
     \hline
     $P_I$ (d) & 5.5 $\pm 0.6$ & 12.3 $\pm 2.0$ & 11.3 $\pm 0.8$ \\
     $\sigma_I$ (d) & 2.7 $\pm 1.4$ & 6.4 $\pm 2.3$ & 5.4 $\pm 1.3$ \\
     $\rho_I$ (\%) & 84 $\pm 12$ & 64 $\pm 12$ & 72 $\pm 14$ \\
     $\rho_{II}$ (\%) & 16 $\pm 12$ & 36 $\pm 12$ & 28 $\pm 14$ \\
     Det. par. & $S/N$ & $w$ & $A$ (mmag) \\
     thres & 1.00 $\pm 0.65$ & 0.25 $\pm 0.08$ & 1.78 $\pm 0.61$ \\
     $D_{\rm KS}$ & 0.058 & 0.034 & 0.025 \\
     $P_{\rm KS}$ (\%) & 51$^c$ & 0.65 & 20 \\
     \hline\hline
     \end{tabular} \\ }
     \small
     \vspace{0.1in}
     {\bf Notes.}\\
     $^a$ $P_I$ and $\sigma_I$ refer to the normal distribution peak and standard deviation for group~I,
        $\rho_I$ and $\rho_{II}$ depict the proportion of stars in groups~I and~II with respect to the entire Sun-like sample,
        \textit{Det. par.} and \textit{thres} give the detectability parameter considered in the simulation
        and the corresponding threshold value obtained in the fit,
        and $D_{\rm KS}$ and $P_{\rm KS}$ are the resulting K-S distance and K-S probability.\\
     $^b$ For the MH$_{\odot}$ sample, the detectability parameter $w$ was simulated from a calibration
        with the amplitude $A$ for an approximate solution.\\
     $^c$ $P_{\rm KS}$ increases with decreasing $D_{\rm KS}$ and decreases with increasing the sample sizes.
Hence, the relatively low number of DMS$_{\odot}$-sample targets produces a high probability even though the MH$_{\odot}$ and RH$_{\odot}$ samples have lower $D_{\rm KS}$ values.
  \label{tabfits}
\end{table}

We then analyze the empirical $P_{\rm rot}$ distribution of each Sun-like subsample
and consider synthetic biased distributions to understand the observational biases.
Because the synthetic samples were produced using models that fit the solar evolution without bias well,
the solar rotation normality can be discussed from a comparison between the actual and modeled distributions.
For an overview of these results,
Table~\ref{tabfits} summarizes the best-fit solutions obtained for the three Sun-like samples considered in this work.
Details are provided in Sects.~\ref{dm13}--\ref{rein}.

\subsection{Evolution across the HR diagram and $P_{\rm rot}$ validation for the CoRoT sample}
\label{corothr}

\begin{figure}
\begin{flushright}
\includegraphics[width=8.98cm]{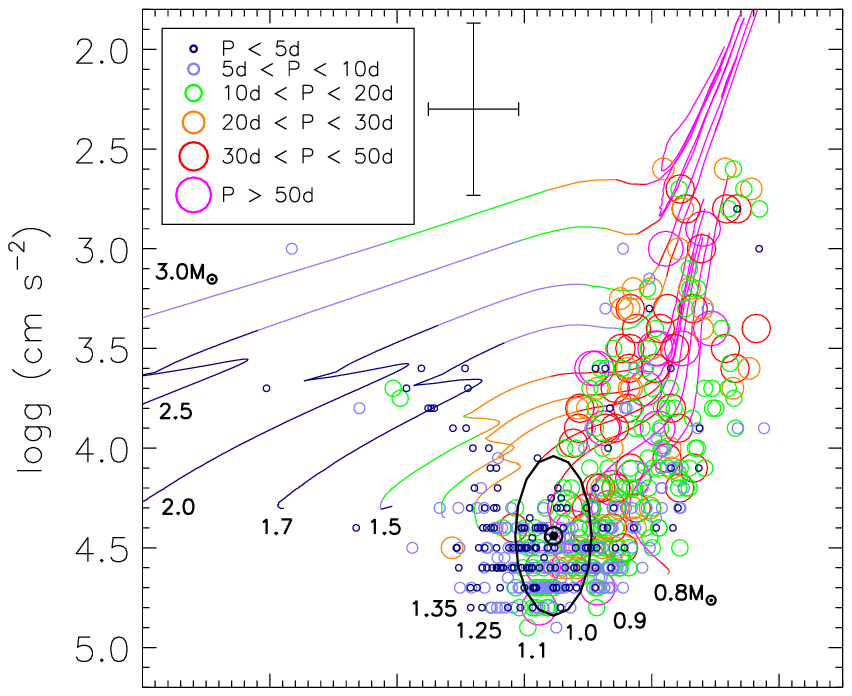}
\includegraphics[width=8.8cm]{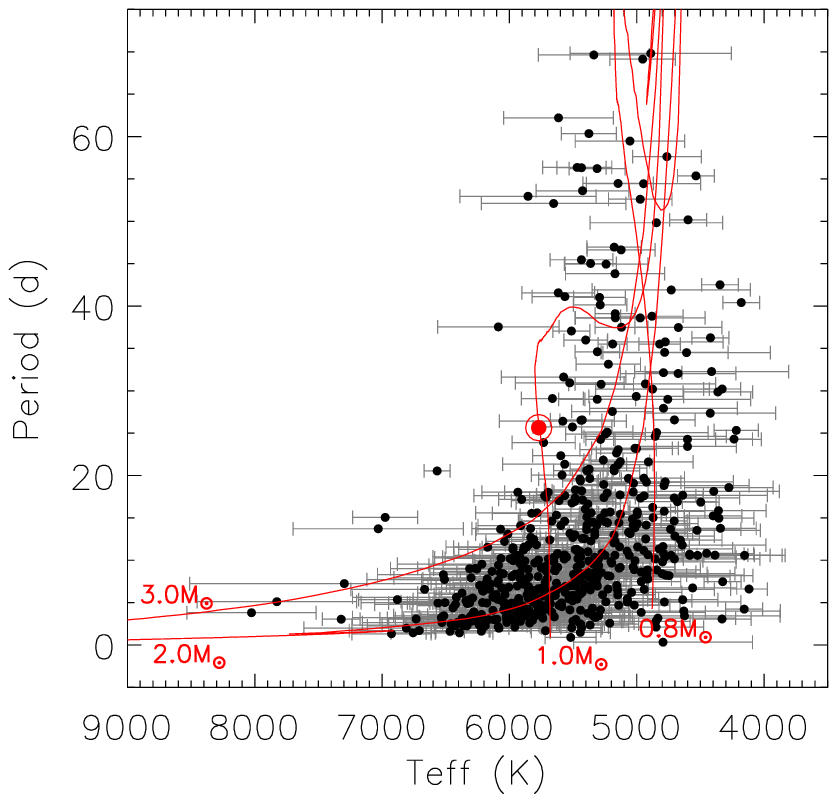}
\caption{
{\it Top panel:} HR diagram for the CoRoT rotating candidates of DM13 combined with the physical parameters of \citet{sar13}, namely, the DMS sample. Circle sizes and colors represent the period intervals shown in the legend. Theoretical tracks from \citet{eks12} are plotted as solid lines for different masses, identified by corresponding labels. The colors represent theoretical rotation periods within the same intervals as in the legend. The typical $T_{\rm eff}$ and $\log g$ uncertainties are illustrated by the error bars. The Sun-like sample is denoted within the ellipsis around the solar location. {\it Bottom panel:} photometric period as a function of effective temperature for the same sample as the top panel. The red solid lines represent theoretical tracks from \citet{eks12} for four different masses. The Sun is denoted in both panels by the solar symbol.
}
\label{hrdiag}
\end{flushright}
\end{figure}

A first important result obtained from the DMS sample is the period distribution across the HR diagram. Figure~\ref{hrdiag} shows this distribution plotted with theoretical tracks from \citet{eks12}. The figure also shows a diagram of periods as a function of $T_{\rm eff}$. This rich combination of photometric periods, spectral parameters, and theoretical predictions encompasses an ample set of cross-checking features for our physical interpretations.

Figure~\ref{hrdiag} (bottom panel) strongly supports the self-consistence between the models considered by \citet{eks12} for rotating stars and the DM13 period values. The total independence of the models and these observations means that the physical ingredients considered in the models of \citet{eks12} are compatible with observations. At the same time, with the new data used here, this result validates the DM13 periods (previously considered as rotational {\it \textup{candidates}}) as being rotational. Although false positives may be present in this sample, they would clearly be a negligible fraction.

Interestingly, the DM13 sample illustrates that this CoRoT subset is mainly distributed around the theoretical track representing the evolving Sun.
This distribution is a result of the selection methods described in DM13, which provided a subsample characterized by well-defined semi-sinusoidal photometric variations (see Sect.~\ref{methods}). This suggests that the semi-sinusoidal signature defined in DM13 is a special characteristic of late-type low-mass stars, with a high occurrence for Sun-like stars at different evolutionary stages. This may have important implications in the study of chromospheric activity and spot behaviors in rotating stars and in their relation to the Sun.
Regarding the temperature-period diagram, the Sun location is sparsely populated,
and many more stars are observed with short periods around the solar temperature.
Biases must be considered when determining whether this represents a trend for the Sun rotating more slowly than Sun-like stars.

\subsection{$P_{\rm rot}$ distribution of the CoRoT sample around the location of the Sun in the HR diagram}
\label{dm13}


\begin{figure}
\begin{center}
\includegraphics[width=8cm]{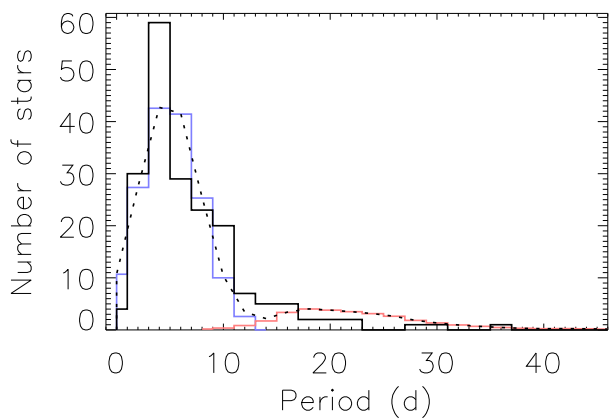}
\caption{
Rotation period distribution of a subset of CoRoT stars with $T_{\rm eff}$ and $\log g$ similar to those of the Sun
 (the DMS$_{\odot}$ sample).
The original sample was obtained from DM13 combined with \citet{sar13} (the DMS sample).
The black histogram shows the actual distribution.
The blue and red histograms and the dotted curve
correspond to a modeled distribution of the actual data where biases were simulated.
Blue depicts a group of young stars (group~I),
red depicts a group of MS and evolved stars (group~II),
and the dotted curve is the sum of these two group distributions.
The latter, which is the full distribution, is the synthetic data considered to fit with observations.
For illustrative purposes, the levels of the modeled distributions were adjusted to make the total number of synthetic stars
similar to that of the actual sample.
}
\label{hist1m_demed}
\end{center}
\end{figure}

Figure~\ref{hist1m_demed} shows the period distribution for a selection of the DMS sample
within an elliptical region around $T_{\rm eff} = 5772 \pm 300$~K and $\log g = 4.44 \pm 0.4$~dex, as depicted in Fig~\ref{hrdiag} (top panel), namely, the DMS$_{\odot}$ sample.
The distribution is asymmetric, with a clear peak around four days and a long tail toward long periods, and it can be interpreted using a
separation into two groups. We do not claim any evidence for a bimodal distribution, but the observations are compatible with it.
Based on this assumption, there is a high occurrence of young stars, corresponding to group~I, compared to group~II.
The reason why considerably fewer stars are observed with an increasing period is explained by a detectability bias.
This bias has its origin in the relations identified by \citet{mes01,mes03}, where the variability amplitude decreases as long as its period increases.
We provide a quantitative analysis of the rotational modulation detectability (related with amplitude)
to understand how actual samples of photometrically rotating stars can be biased.
Thus, our method is an important contribution to analyze $P_{\rm rot}$ by comparing actual biased samples with theory.
Other authors \citep[e.g.,][]{aig15} have recently investigated methods to provide better constraints on the rotational modulation detectability.

For the synthetic sample, we applied a normal distribution to generate the rotation periods of group~I, as discussed in Sect.~\ref{pms}.
For this sample, the synthetic distribution fit observations
well by centering at approximately $5.5 \pm 0.6$~d with a standard deviation of approximately $2.7 \pm 1.4$~d (see~Table~\ref{tabfits}).
These values are compatible with those computed in Sect.~\ref{pms} for a compiled sample of PMS stars.
Exponential decays were also tested, but they provided similar fits when considering the full synthetic distribution.
For group~II, the rotation periods were obtained from interpolating theoretical grids from \citet{eks12}, as explained in Sect.~\ref{methods}.
The proportions $\rho_I$ and $\rho_{II}$ segregated by the 1~Gyr isochrone were 84\% and 16\%, respectively, which fits the observed distribution without any additional adjustment.
Thus, the full superposition of groups~I and~II in Fig.~\ref{hist1m_demed} corresponds with the observed distribution within the limitations of the CoRoT data.
We cannot infer any conclusion about the solar rotation normality from this CoRoT sample
because the data are too sparse around the solar period.
However, the compatibility between the synthetic and observed distributions
illustrates that the synthetic sample was generated based on reasonable assumptions
and could be a start to validate our approach.
In addition, this result provides a basic explanation for the apparent
abnormality of the solar rotation compared to Sun-like stars, as described in Sect.~\ref{corothr}.
The Sun-like samples obtained in this work cover very many young stars.
This means that the many targets that rotate faster than the Sun in our Sun-like samples are most likely young objects.

The CoRoT sample analyzed above was limited by a relatively low number of stars
in comparison to the {\em Kepler} samples because of at least two reasons.
First, based on Fig.~\ref{magnoise}, CoRoT covers a $V_{\rm mag}$ range of $\sim$12--18~mag
and noise levels of $\sim$600--$10,\!000$~ppm,
while for {\em Kepler} the $V_{\rm mag}$ range is of $\sim$8--18~mag
and the noise levels of $\sim$30--5000~ppm.
As such, {\em Kepler} has a higher photometric sensitivity
and can observe brighter stars than CoRoT, which means that \textit{Kepler} has more potential
to detect micro-variability signals (such as the rotational modulation) for many stars.
Second, the catalog of physical parameters obtained by~\citet{sar13} (which we combined with the
catalog of DM13) has relatively few stars ($6,\!832$ sources),
whereas the catalog of~\citet{hub14} provides physical parameters for almost all {\em Kepler} targets ($196,\!468$ stars).
{\em Kepler} therefore has considerably more data for $P_{\rm rot}$ measurements and for physical
parameters and hence provides more reliable results, as presented below in Sects.~\ref{mcq14} and~\ref{rein}.

\subsection{Testing a recent {\em Kepler} sample}
\label{mcq14}

\begin{figure}
\begin{center}
\includegraphics[width=8cm]{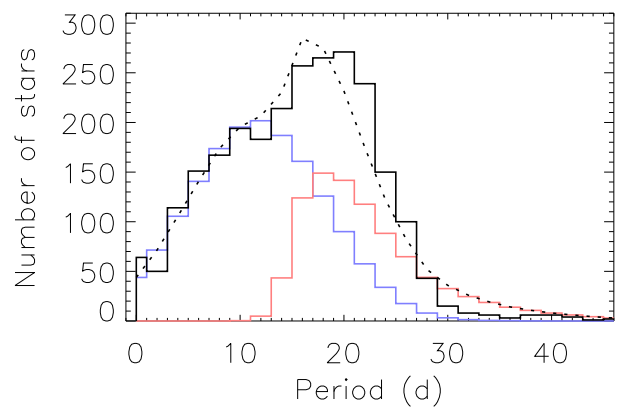}
\caption{
Rotation period distribution of a newer subset of {\em Kepler} stars with $T_{\rm eff}$ and $\log g$ similar to those of the Sun
(the MH$_{\odot}$ sample).
The original sample was obtained from~\citet{mcq14} combined with \citet{hub14}.
The symbols are the same as in Fig.~\ref{hist1m_demed}.
}
\label{hist1m_mcq14}
\end{center}
\end{figure}

As a test, Fig.~\ref{hist1m_mcq14} shows the period distribution of a Sun-like {\em Kepler} sample of rotating stars obtained by \citet{mcq14}
combined with the physical parameters of \citet{hub14}, namely the MH$_{\odot}$ sample.
This stellar set was selected within an elliptical region around the solar location in the HR-diagram,
as illustrated in Fig.~\ref{hessparent} (bottom panel).
This is considerably more populated than the CoRoT sample and could be selected closer to the solar metallicity,
specifically $-0.2$~dex~$< [Fe/H] < 0.2$~dex.
Therefore, it may allow for a more refined analysis of the rotation period distribution based on the synthetic calculations.
However, for this case, simulating the detectability parameter $w$ properly was particularly challenging;
this limits our interpretaion of the MH$_{\odot}$ sample.
We consider this case for an overview of the results and to discuss the importance of simulating
the detectability parameter well. For the RH$_{\odot}$ sample (see Sect.~\ref{rein}), we were able to properly simulate a detectability cutoff ,
and we obtained an optimal fit between simulation and observations.

Overall, the $P_{\rm rot}$ distribution also has a bimodal shape and can be decomposed into groups~I and~II,
as performed in the DMS$_{\odot}$ sample.
In this test, the proportions $\rho_I$ and $\rho_{II}$
initially set up by separating the parent sample into the regions below and above
the 1~Gyr isochrone (which provided $\rho_I = 81$\% and $\rho_{II} = 19$\%)
did not fit the observations well.
A better fit was obtained by adjusting these proportions to 64\% and 36\%, respectively.
This handle is a fair adjustment related with the period-amplitude distribution and with the line that separates
the two groups (see~Sect.~\ref{simul}).
The simulation of group~I was simply defined as a normal distribution, which agreed with observations by peaking at approximately $12.3 \pm 2.0$~d with a standard deviation of approximately $6.4 \pm 2.3$~d.
Hence, this distribution is centered at longer periods and has a broader spread than in the CoRoT sample.
This difference may be related to the distinct galactic regions and fields of view (FOV) of the CoRoT and {\em Kepler} stellar sets,
as analyzed in Sect.~\ref{rein}.
Exponential decays also provided similar fits, which did not improve the results.

Regarding the superposition of group~I with group~II,
the decay of the synthetic distribution to the right of the second peak reaches from slightly steeper to smoother than in the observed distribution.
As a result, the synthetic sample presents a small excess of stars for periods somewhat above the solar value
in comparison with observations.
The discrepancy might be reduced if practical difficulties of measuring long periods
were implemented in our approach.
To check for this possibility, we tested convolving the synthetic distribution with a linear decay that starts in a 100\% probability at $P_{\rm rot} = 0$ and decreases with a coefficient that was set as a free parameter.
This function is compatible with the recovery fraction as a function of the number of observed variability
cycles, as analyzed in Sect.~2.2.2 of~DM13.
In this test an excess remained at long periods, so this contribution could not be sufficient to adjust the excess.
Therefore, such an alternative ingredient would not produce a degeneracy in the fit solution
and the discrepancy is likely due to our limitations in simulating $w$.
This conclusion is discussed further in Sect.~\ref{rein}, where the simulation {of $R_{\rm var}$} was
easily approximated to the $A$ numerical values.

The discrepancy occurred in Fig.~\ref{hist1m_mcq14} illustrates the general effect observed in the $P_{\rm rot}$ distribution
if a proper detectability cutoff is not performed in the synthetic samples.
This effect is furthermore noticeable if the proportions $\rho_I$ and $\rho_{II}$
are adjusted as free parameters without performing a detectability cutoff (see Sect.~\ref{simul}).
Overall, $\rho_I$ and $\rho_{II}$ can balance the peak levels of each group,
whereas a detectability cutoff produces a fine-tuning of the distribution slope at long periods.
In particular, the detectability thresholds obtained here as free parameters were substantially plausible (see Table~\ref{tabfits}):
$S/N = 1.00$ and $w = 0.25$ perfectly match the cutoffs used by DM13
and \citet{mcq14}, respectively, whereas $A = 1.78$~mmag is compatible with the threshold of $R_{\rm var} = 3\permil$
defined by \citet{rei13}.
Despite the discrepancy in Fig.~\ref{hist1m_mcq14}, the K-S test provided a non-null probability,
and the synthetic sample fit observations reasonably well.
Thus, this fact indicates the normality of the solar $P_{\rm rot}$ compared to Sun-like stars
based on the MH$_{\odot}$ sample.

\subsection{In-depth study of another {\em Kepler} sample}
\label{rein}

\begin{figure}
\begin{center}
\includegraphics[width=8cm]{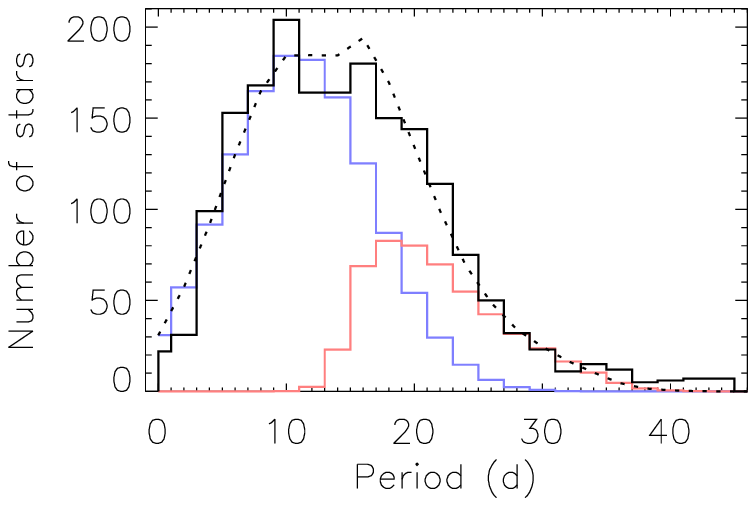}
\caption{
Rotation period distribution of a subset of {\em Kepler} stars with $T_{\rm eff}$ and $\log g$ similar to those of the Sun
(namely, the RH$_{\odot}$ sample).
The original sample was obtained from~\citet{rei13} combined with \citet{hub14}.
The symbols are the same as in Fig.~\ref{hist1m_demed}.
}
\label{hist1m_rei13}
\end{center}
\end{figure}

Figure~\ref{hist1m_rei13} presents the rotation period distribution of a Sun-like subset of sources of~\citet{rei13} combined with~\citet{hub14}, namely, the RH$_{\odot}$ sample.
This is a highly populated stellar set, similar to the MH$_{\odot}$ sample,
also selected within the elliptical region depicted in Fig.~\ref{hessparent} (bottom panel) and
within $-0.2$~dex~$< [Fe/H] < 0.2$~dex.
In addition, the detectability parameter of the RH$_{\odot}$ sample, $R_{\rm var}$, can be easily simulated
because its numerical value is highly similar to that of the variability amplitude $A$.
Therefore, this sample allows for the most detailed and complete analysis of the $P_{\rm rot}$ distribution in the present work.

The $P_{\rm rot}$ distribution of the RH$_{\odot}$ sample is noticeably different than that of the MH$_{\odot}$ sample.
One reason is certainly the different time spans of the LCs
(the RH$_{\odot}$ sample uses only Q3 data, the MH$_{\odot}$ sample uses Q3-Q14 data),
which facilitates long-period detection in the MH$_{\odot}$ sample.
Despite the difference, the distribution is also bimodal and can be decomposed into groups~I and~II,
as performed for the samples analyzed above.
In the synthetic sample, when groups~I and~II were initially decomposed by the regions below and above
the 1~Gyr isochrone (giving $\rho_I$ and $\rho_{II}$ of 81\% and 19\%, respectively), the $P_{\rm rot}$ distribution fit the observations reasonably well.
However, the best fit was obtained by adjusting $\rho_I$ and $\rho_{II}$ to 72\% and 28\%, respectively.
As suggested in Sect.~\ref{pms}, a normal distribution was defined for group~I, which agreed with observations by peaking at approximately $11.3 \pm 0.8$~d with a standard deviation of approximately $5.4 \pm 1.3$~d.
Hence, this distribution is compatible with the RH$_{\odot}$ sample.
Exponential decays also provided acceptable fits and produced similar behaviors, which did not improve the results.

For the entire $P_{\rm rot}$ distribution of this sample, a noteworthy fit was obtained
because the simulation was able to properly reproduce the bias caused by the detectability parameter $R_{\rm var}$.
Indeed, we verified that for the RH$_{\odot}$ sample, if $S/N$ was used as the detectability parameter instead of $A$,
a discrepancy would occur between the simulation and observations in the distribution decay around the solar period,
similar to the problem of the MH$_{\odot}$ sample.
Therefore, an appropriate simulation of the detectability parameter certainly contributes to an optimal fit with observations.
Because the synthetic distribution has the theoretical grid of \citet{eks12} as a basic ingredient,
which unbiasedly fit the Sun and Sun-like stars,
the good fit indicates that the solar $P_{\rm rot}$ is normal with regard to Sun-like stars.

Compared to the other CoRoT and {\em Kepler} samples, the period distribution of group~I is compatible with the RH$_{\odot}$ sample, and centered around a longer period and with a broader spread than in the DMS$_{\odot}$ sample.
This supports the possible relation of the group~I distribution with the stellar population,
as mentioned in Sect.~\ref{mcq14}.
Indeed, the CoRoT fields lie toward the center and anti-center of the Galaxy, covering
very many stars located near the Galactic plane.
This region is expected to be more populated by young stars than at high Galactic latitudes \citep[e.g.,][]{mig13}.
The CoRoT fields are also small, covering together a region of $\sim$8 square degrees.
In contrast, the {\em Kepler} field is considerably larger ($\sim$105 square degrees) and covers
a broad range of Galactic latitudes \citep{mig13,gir15}.
These facts suggest a higher
proportion and a narrower distribution of young stars in the CoRoT sample than in the {\em Kepler}
sample. To test this hypothesis, we used the TRILEGAL code
to obtain a sampling of the CoRoT and {\em Kepler} fields,
covering nearly the same regions of their actual
coordinates. The limiting magnitude was chosen
to be $V = 16$~mag for the CoRoT fields and $K_p = 17$~mag
for the {\em Kepler} field. Then we restricted the
$T_{\rm eff}$, $log g$, and $[M/H]$ (assumed to be $\simeq$$[Fe/H]$) ranges
according to our own selection of Sun-like stars.
Based on this test, the age distribution of the CoRot and {\em Kepler} samples
agrees well with the $P_{\rm rot}$ distributions presented here, where the CoRoT fields include a greater proportion of young stars than older stars.
In the {\em Kepler} field simulation, the simulation is much broader,
with a considerably fewer young stars than in the CoRoT sample.

Despite this difference, either the CoRoT or {\em Kepler} fields comprise a high proportion of young objects. Overall, 64--84\% of the Sun-like samples analyzed here are younger than $\sim$1~Gyr.
This result (restricted to Sun-like stars) is reasonable if compared with a related estimate from \citet{mat14}, valid for the whole sample of {\em Kepler} sources, based on a relatively simple stellar rotation model.
According to these authors, $\sim$95\% of the whole sample of {\em Kepler} field stars with $P_{\rm rot}$ measurements
are younger than $\sim$4~Gyr.

In general, our study shows that the CoRoT and {\em Kepler} data can be well interpreted in terms of two populations, one young (group~I), one older (group~II), and with a different ratio between them. The proportions $\rho_I$ and $\rho_{II}$ for the CoRoT and {\em Kepler} samples considered here can be  well interpreted  by simulating the experiment sensitivity and biases.
At least on the basis of the present data, there is no clear reason to believe that the bimodality reported by \citet{mcq13a,mcq14} can be explained by selection biases. As such, we support the suggestion of these authors, with some addition. The bimodality is possibly explained by a ``natural bias'' (i.e., not dependent on measurement sensitivity, but related to evolutionary duration) considering that two stellar populations experience evolutionary phases that have very different durations. These populations are, in fact, the groups~I and~II defined here. Therefore, the detectability bias is not the reason of the bimodality, although it influences the proportions $\rho_I$ and $\rho_{II}$.

Of course, the normal distribution adopted here for group~I is an approximation that may have a number of implicit biases.
In contrast, the distribution of group~II is based on a more elaborate set of theoretical models and assumptions,
including the theoretical grids provided by \citet{eks12}, modeled parent distributions, the empirical period-amplitude relations by~\citet{mes01,mes03}, the selection by $S/N$, $A,$ or $w$ (which mimic detectability biases of the rotational modulation), and final filtering by Sun-like characteristics, all described in Sect.~\ref{simul}.
Hence, the set of ingredients we presented allows us to study solar rotation by comparing actual and biased observations with theoretical models, within a precision that depends on the current data and model limitations.
More refined studies can be developed in the future with this method.
In particular, a potential improvement of the current results would be possible
if spot-induced variability amplitudes were included in
the stellar evolution codes with rotation.

%

\section{Conclusions}

We presented for the first time
a quantified method with which the empirical distribution of $P_{\rm rot}$ for Sun-like stars
can be explained by considering the detectability of rotational modulation, which is related to the variability amplitude.
The interpretations were based on a comparison between synthetic and empirical subsamples of CoRoT and {\em Kepler} field stars with physical parameters similar to those of the Sun.

For this purpose, we combined the public data of rotation periods with physical parameters for three stellar samples.
The first sample was obtained from a CoRoT catalog with period measurements given by~DM13, combined with spectroscopic physical parameters $\log g$ and $T_{\rm eff}$ determined by \citet{sar13}, namely, the DMS~sample.
The selection of Sun-like stars was obtained within an elliptical region around the solar location in the HR~diagram. For CoRoT stars, the subsample was enclosed by $T_{\rm eff} = 5772 \pm 300$~K and $\log g = 4.44 \pm 0.4$~dex, namely, the DMS$_{\odot}$ sample.
The two other samples were selected from {\em Kepler} catalogs of rotation periods provided by \citet{rei13} and \citet{mcq14}, each combined with an improved compilation of physical parameters given by \citet{hub14}.
The Sun-like subsamples were selected within $-0.2$~dex~$< [Fe/H] < 0.2$~dex and within an ellipsis described by $T_{\rm eff} = 5772 \pm 170$~K and $\log g = 4.44 \pm 0.2$~dex, namely, the RH$_{\odot}$ and MH$_{\odot}$ samples.

To explain the observational biases, synthetic samples were generated based on the parent samples of the CoRoT and {\em Kepler} fields.
The synthetic stars were separated into two groups, I and II, which are those located below and above the 1~Gyr isochrone in the HR~diagram, respectively. These represent a group of young stars and a group composed of MS and evolved stars, respectively.
For group~I, synthetic rotation periods were randomly generated following two possible distributions compatible with PMS Sun-like stars: normal or an exponential decay from the peak.
For group~II, theoretical periods from~\citet{eks12} were set up according to the HR~diagram location of the synthetic stars.
Next, period-amplitude relations from \citet{mes01,mes03} were used to aggregate amplitude values to the synthetic samples, and
random noise levels were added to these samples.
Then, the main bias related to the detectability of the rotational modulation was applied to the synthetic stars by selecting them with a threshold in the variability $S/N$, $A,$ or $w$ (from \citealt{mcq14}, which was particularly difficult to simulate) according to each observed sample.
Finally, Sun-like subsamples were obtained by selecting the targets located within the ellipses described around the solar location in the HR~diagram from the synthetic samples.
The ellipses were defined within the same regions as for each one of the three observational subsamples of Sun-like stars considered here, namely, the DMS$_{\odot}$, RH$_{\odot}$, and MH$_{\odot}$ samples.

The distribution in the HR diagram of the DMS sample has a large number of stars with low masses that are particularly similar to that of the Sun. This sample was obtained by identifying a well-defined semi-sinusoidal signature in the light curves without knowledge of their physical parameters, apart from photometric magnitudes and colors. This sample indicated that the semi-sinusoidal signature, defined by DM13, provides a good selection of low-mass spotted rotating stars, with a high occurrence for stars with masses of~$\sim$1~M$_{\odot}$, which may have important implications in the relation between rotation and chromospheric activity for evolving Sun-like stars.

The rotation period distribution of the DMS$_{\odot}$ sample is not clearly bimodal, but it fits our bimodal model, thus showing
a high occurrence of young stars.
The proportions of stars belonging to groups~I and~II with respect to the total number of objects
in the synthetic sample, namely, $\rho_I$ and $\rho_{II}$, are
of 84\% and 16\%, respectively. These proportions fit the observations
when simply separated by the 1~Gyr isochrone without any complementary adjustment.
In addition, the full distribution of the synthetic sample, that
is, the sum of groups~I and~II,
exhibited a close fit with the observed distribution.
This shows that separating the two groups by using the 1~Gyr isochrone is reasonable.
Because of the relatively low number of objects in the DMS$_{\odot}$ sample (175 targets),
we cannot make conclusions regarding the solar rotation normality with respect to its Sun-like counterparts.
Therefore, this result helps to validate the methods used to produce our biased synthetic samples.
Further in-depth analyses can be performed from the {\em Kepler} data.

For the MH$_{\odot}$ sample, the period distribution is more seemingly
bimodal and has considerably more stars ($2,\!525$ targets) than the DMS$_{\odot}$ sample.
For this case, the proportions $\rho_I$ and $\rho_{II}$ originally did not fit the observations well
and was adjusted to final proportions of 64\% and 36\%, respectively.
With this adjustment, the modeled distribution fit the observations
well,
although it exhibited a noticeable discrepancy in the distribution decay that traverses the solar rotation period.
However, near the solar period, the synthetic sample fits the observations well,
which suggests that the solar rotation period is normal with respect to Sun-like stars.

The period distribution of the RH$_{\odot}$ sample also shows a bimodality and has a high number of stars ($1,836$ targets),
similar to the MH$_{\odot}$ sample.
The proportions $\rho_I$ and $\rho_{II}$ originally fit the observations well,
although the best fit was obtained by adjusting these parameters to the final values of 72\% and 28\%, respectively.
With this adjustment, the full period distribution of the synthetic sample provided an extremely close fit with observations.
This fact strongly supports that the solar $P_{\rm rot}$ is normal with respect to Sun-like stars.

Overall, the main conclusions of this work can be summarized as follows:
\begin{enumerate}[i.]
\item The synthetic samples were able to explain the observed $P_{\rm rot}$ distributions
by resolving them into two different populations, group~I for young stars
and group~II for main-sequence and evolved stars, which generally produce a bimodal arrangement.
\item The central peak and the width of the group~I $P_{\rm rot}$ distribution
may be related with the stellar population within a certain FOV.
\item Reasonable to optimal fits of the synthetic samples with the observed data were obtained
especially by adjusting the bimodal peak levels with $\rho_I$ and $\rho_{II}$ and
by refining the distribution slope at long periods with an appropriate detectability threshold for each Sun-like sample.
\item Several tests indicated that the parameters considered in these fits (see Table~\ref{tabfits})
provide a simple description of the $P_{\rm rot}$ distributions without degeneracy.
\item A detectability threshold is mandatory to properly fit synthetic with observed data for the CoRoT sample,
while it improves the fits for the {\em Kepler} samples, but with a weaker effect.
This suggests the {\em Kepler} samples are less biased by the variability amplitude than the CoRoT sample.
\item Best fits suggest a high number (64--84\%) of young objects in the Sun-like samples,
which explains the short $P_{\rm rot}$ values of Sun-like stars on average compared to the solar value.
\item The global agreement between the synthetic and observed samples suggests the normality of the solar $P_{\rm rot}$,
at least within the current data accuracy and model limitations.
\end{enumerate}
Therefore, the method presented in this work
allowed us to constrain observations on theory by considering biased samples with $P_{\rm rot}$ measurements
and was particularly useful to analyze field Sun-like stars.
In addition, the following perspectives can improve our results:
\begin{enumerate}[i.]
\item A substantial improvement of the synthetic samples will be possible once
stellar evolution models can predict the photometric amplitude variations produced by rotating star spots.
Such a theoretical prediction needs a description of the magnetic activity at the surface of stars,
and the interpretation of the observed relation between $P_{\rm rot}$ and $A$ is one of the observational facts that can help
in this subject.
\item Observed data with more accurate physical parameters will provide a more refined
selection of Sun-like stars and a more reliable separation of groups~I and~II.
Accordingly, future spectroscopic observations, such as those to be collected with the
Gaia-ESO Survey\footnote{\tt http://www.gaia-eso.eu/} \citep{gil12}, will allow a better verification of our results.
\end{enumerate}

%

\begin{acknowledgements}

This paper includes data collected by the Kepler mission. Funding for the Kepler mission is provided by the NASA Science Mission directorate.

Research activities of the Observational Astronomy Stellar Board at the Federal University of Rio Grande do Norte are supported by continuous grants from the
brazilian agencies CNPq and FAPERN and by the INCT-INEspa\c{c}o.
I.C.L. acknowledges post-doctoral PNPD/CNPq and post-doctoral PDE/CNPq fellowships. I.C.L. also thanks ESO/Garching for hospitality. C.E.F.L. acknowledges a post-doctoral PDJ/INCT-INEspa\c{c}o/CNPq fellowship.
L.P. acknowledges a distinguished visitor PVE/CNPq appointment at the DFTE/UFRN. L.P. also thanks DFTE/UFRN for hospitality and support of the INCT-INEspa\c{c}o.
V.N. acknowledges A BJT/CNPq fellowship. L.L.A.O. and D.F.S. acknowledge MSc fellowships of the CNPq.
A.A.R.V.  acknowledges support from FONDECYT postdoctoral grants \#3140575, Ministry for the Economy, Development, and Tourism's Programa Iniciativa Cient\'{i}fica Milenio through grant IC\,210009, awarded to the Millennium Institute of Astrophysics (MAS), CAPES grant PNPD/2011-Institucional, and the CNPq Brazilian agency.
The authors thank the anonymous referee for very valuable comments and suggestions.
We warmly thank F. Gallet for kindly providing us his catalog of rotation periods of PMS Sun-like stars, which was used in this work.
The authors thank the CoRoT and {\em Kepler} staff for the development, operation, maintenance, and success of the mission. This work used the VizieR Catalogue Service operated at the CDS, Strasbourg, France.

\end{acknowledgements}


\end{document}